\documentclass[a4paper,fleqn,usenatbib]{mnras}

\usepackage{newtxtext,newtxmath}

\usepackage[T1]{fontenc}
\usepackage{ae,aecompl}

\usepackage{graphicx}	
\usepackage{amsmath}	
\usepackage{amssymb}	
\usepackage{siunitx}
\usepackage[flushleft]{threeparttable}

\title[XMMU J181227.8--181234]{XMMU J181227.8--181234: a new ultracompact X-ray binary candidate}

\author[A. J. Goodwin et al.]{
A. J. Goodwin,$^{1,2}$\thanks{E-mail: adelle.goodwin@monash.edu}
D. K. Galloway,$^{1,2}$
J. J. M. in 't Zand,$^{3}$
E. Kuulkers,$^{4}$
A. Bilous$^{5}$
\newauthor
and L. Keek$^{6}$
\\
$^{1}$School of Physics and Astronomy, Monash University, Clayton, Victoria, Austalia, 3800\\
$^{2}$also Monash Centre for Astrophysics\\
$^{3}$SRON Netherlands Institute for Space Research, Sorbonnelaan 2, 3584 CA Utrecht, The Netherlands\\
$^{4}$ ESA/ESTEC, Keplerlaan 1, 2201, AZ Noordwijk, The Netherlands\\
$^{5}$ Anton Pannekoek Institute for Astronomy, University of Amsterdam, Science Park 904, 1098 XH Amsterdam, The Netherlands \\
$^{6}$ Department of Astronomy, University of Maryland, College Park, MD 20742, USA \\
}

\date{Accepted 2019 April 15. Received 2019 April 13; in original form 2018 December 20}

\pubyear{2019}

\begin{document}
\label{firstpage}
\pagerange{\pageref{firstpage}--\pageref{lastpage}}
\maketitle

\begin{abstract}
We report the discovery of Type I (thermonuclear) X-ray bursts from the transient source XMMU J181227.8--181234 = XTE J1812--182. We found 7 X-ray bursts in \textit{Rossi X-ray Timing Explorer} observations during the 2008 outburst, confirming the source as a neutron star low mass X-ray binary. Based on the measured burst fluence and the average recurrence time of 1.4$^{+0.9}_{-0.5}$  ~hr, we deduce that the source is accreting almost pure helium ($X \leq 0.1$) fuel. Two bursts occurred just 18 minutes apart; the first short waiting time bursts observed in a source accreting hydrogen-poor fuel. Taking into consideration the effects on the burst and persistent flux due to the inferred system inclination of $30\pm{10}$\textdegree, we estimate the distance to be $14\pm{2}$~kpc, where we report the statistical uncertainty but note that there could be up to $20\%$ variation in the distance due to systematic effects discussed in the paper. The corresponding maximum accretion rate is $0.30\pm0.05$ times the Eddington limit. Based on the low hydrogen content of the accreted fuel and the short average recurrence time, we classify the source as a transient ultracompact low-mass X-ray binary.
\end{abstract}




\section{Introduction}

Low mass X-ray binaries are neutron stars or black holes accreting from a companion star in a close binary orbit \citep[e.g.][]{white1995}. As hydrogen/helium accretes onto a neutron star (NS), the hydrogen may burn steadily via the (hot) CNO cycle. If the accreting material condenses and heats enough, reaching a high enough pressure to cause a thermonuclear runaway, a Type I X-ray burst is produced  \citep[see][for comprehensive reviews]{lewin1993,galloway2017}. There are two types of emission that can be observed: the usually fainter, persistent emission that is produced by the accretion process onto the NS, and the short, bright flares that are X-ray bursts.

Observations of Type I X-ray bursts can constrain the composition of fuel via the nuclear energy generation of the bursts, the recurrence time of the bursts and the accretion rate at which the material is piling onto the NS. The distance and inclination of the source can be constrained if photospheric radius expansion is observed \citep[e.g][]{galloway2008}. The ratio of gravitational energy to nuclear burning energy is higher for sources that burn more helium during a burst, as helium burns via the triple alpha process, producing less energy per nucleon than hydrogen, which burns via the hot CNO cycle and $rp$ process \citep{wallace1981}. Helium-fueled bursts are also observed to have faster rise times than hydrogen bursts \citep[e.g.][]{galloway2008}. The composition of the accreted fuel gives insight into the burning processes that are occurring in the system as well as the evolutionary history of the binary system. For example, a low mass binary with a companion star that donates almost pure helium to the NS must be significantly evolved, as a young, low mass star does not accumulate helium in its core on short timescales.

 XMMU J181227.8--181234 = XTE J1812--182 was first discovered on March 20, 2003 by \textit{XMM-Newton} when it went into outburst, as reported by \citet{cackett2006}. Initially it was only detected with XMM, and not with the \textit{Rossi X-ray Timing Explorer (\textit{RXTE}) All Sky Monitor} (ASM). However, upon reprocessing of the \textit{RXTE} data, it was detected. No pulsations or X-ray bursts were originally detected. \citet{cackett2006} found a very high absorption along the line of sight to the source ($N_{\rm H}$ of 12.8$\pm0.3\times$10$^{22}\,$cm$^{-2}$). They could not conclusively identify if the source was a high mass X-ray binary characteristic of the color and absorption of the source, or low mass X-ray binary characteristic of the steep power law index that fit its spectrum.

On 2008 August 21 \textit{RXTE} detected XMMU J181227.8--181234 in outburst again, in proportional counter array (PCA) scans of the Galactic ridge region \citep{Markwardt2008}. Twenty-six follow-up observations were performed with the PCA between MJD 54699 and 54758 (August 21 and October 19, 2008).

In this paper, we report on the discovery and analysis of bursts of XMMU J181227.8--181234 in these 2008 follow-up observations. An initial report was given by \citet{ouratel2017}. Throughout this paper we report the 1-$\sigma$ confidence intervals when uncertainties are presented, unless otherwise noted. In Section \ref{sec:data} we describe the data extraction and telescope observations; in Section \ref{sec:analysis} we analyse the burst and persistent emission and deduce the source properties, providing a full treatment of the anisotropy factors and thus inclination of the accretion disk. Finally, in Section \ref{sec:classification} we conclude by classifying the source as a transient ultracompact low mass X-ray binary.

\section{Data Reduction}\label{sec:data}

We analysed archival \textit{RXTE} observations of XMMU 181227.8--181234 from its 2008 outburst \citep{cackett2006,Markwardt2008,Torres2008}. We used archival \textit{RXTE} ASM and Proportional Counter Array (PCA) data available via the Multi-Instrument Burst Archive (MINBAR\footnote{{\tt https://burst.sci.monash.edu/minbar/}}) and in the publicly available online \textit{RXTE} archive. The PCA \citep{jahoda2006} consists of five proportional counter units (PCUs), each with approximately 1600 cm$^2$ of photon-sensitive area, and a bandpass of 2--60 keV at a spectral resolution of about 20\% full width at half maximum (FWHM). The PCUs are co-aligned with a collimator delimiting the field of view to 1\textdegree \,(FWHM).

There were a total of 26 dedicated PCA observations between MJD 54699 and 54758, with a total exposure time of 45 ks that the source is visible. There were a further 49 PCA bulge scan observations between MJD
54420 and 55890 during which the source was in quiescence. There were 4043 archival ASM measurements covering the 2--10 keV energy band between MJD 50137 and 55825. 
For the persistent flux observations in Section \ref{sec:longtermflux} we also used \textit{Swift} Burst Alert Telescope (BAT) daily average light curve data from the online publicly available archive \citep{krimm2008}. There were 4049 daily BAT observations between MJD 53415 and 58015 in the 15--50 keV energy band. Finally, we used \textit{BeppoSAX} Wide Field Camera \citep[WFC;][]{jager1997} data from MINBAR in the 2--28 keV energy band. There is 4.5 Msec of WFC data between MJD 50310 and 52380 (1996--2002).

\subsection{Position and History}

\citet{Markwardt2008} reported the \textit{RXTE} detection of a transient source in PCA scans of the Galactic ridge region in an observation on 2008 August 21. They determined the position of this source in a follow up scanning observation to be (RA,Dec) = 273.117, -18.26 (J2000) with an error radius of 2.1$\,$arcmin (95$\%$ confidence), making it likely to be renewed activity of XMMU 181227.8--181234. This position translates to Galactic coordinates $(l^{\rm II}, b^{\rm II})=(12\fdg 3577, +0\fdg 0337)$. The bursting transient source XTE J1810-189 is 0.996$\,$degrees away from this location \citep{krimm2008}, just within the PCA field of view. However, the last significant detection of XTE J1810-189 with the PCA during the Galactic bulge scans was on 2008 June 10, when it concluded its most recent outburst \citep{weng2015}. Since no activity of XTE J1810-189 was detected for the three months following, up to the time of the bursts we detected, we can confidently exclude this source as the source of the detected photons and attribute them to XMMU J181227.8--181234 = XTE J1812--182.

\citet{cackett2006} noticed that XMMU J181227.8--181234 lies $\sim$40\arcsec\ from the 95$\%$ confidence-level 2\fdg1 by 1\farcm8 error box of the source 1H1812--182, which was detected in 1977 during the Large Area Sky Survey Experiment with the HEAO-1 Satellite \citep{woods1984}. They concluded that although the centroid position of 1H1812--182 is $\sim$0.7\textdegree\, from that of XMMU J181227.8--181234, the large extended error box makes it possible that the HEAO 1 source is a detection of XMMU J181227.8--181234 during a previous outburst. \citet{woods1984} reported that 1H1812--182 was detected with a flux of 0.1470 $\pm$ 0.0025 counts cm$^{-2}$ s$^{-1}$ in the 0.5--25 keV band, corresponding to an unabsorbed flux of approximately 9.7 $\times$ 10$^{-10}$ ergs cm$^{-2}$ s$^{-1}$ and absorbed flux of approximately 4.0 $\times$ 10$^{-10}$ ergs cm$^{-2}$ s$^{-1}$ in the 2--25 keV range, assuming a power law persistent spectrum with parameters in Table \ref{tab:spectrafits}. We measured a range of (0.079--1.551) $\times$ 10$^{-9}$ ergs cm$^{-2}$ s$^{-1}$ in the 2--25 keV band with the \textit{RXTE} PCA telescope (see below), which means the flux of 1H1812--182 is within the ranges observed for XMMU J181227.8--181234. Interestingly, \citet{Fleischman1985} reported the observation of a single X-ray burst in August 1977 from either GX 13+1 or 1H1812-182, with the Horizontal Tube Detector aboard SAS-3. The detector had a 3.4\textdegree\, diameter so the author could not discern if the X-ray burst came from GX 13+1 or 1H1812-182. \citet{Matsuba1995} reported the observations of bright bursts from GX 13+1, with similar characteristics to the burst observed by \citet{Fleischman1985}, making it plausible this X-ray burst came from GX 13+1. However, there is a chance this burst came from 1H1812-182 = XMMU J181227.8--181234, or, in the case they are not the same source, this X-ray burst could even have come from XMMU J181227.8--181234, before it was a known source. 

\section{Analysis and Results}\label{sec:analysis}

\subsection{Light Curve and Persistent Spectrum}

\begin{figure}
	\includegraphics[width=\columnwidth]{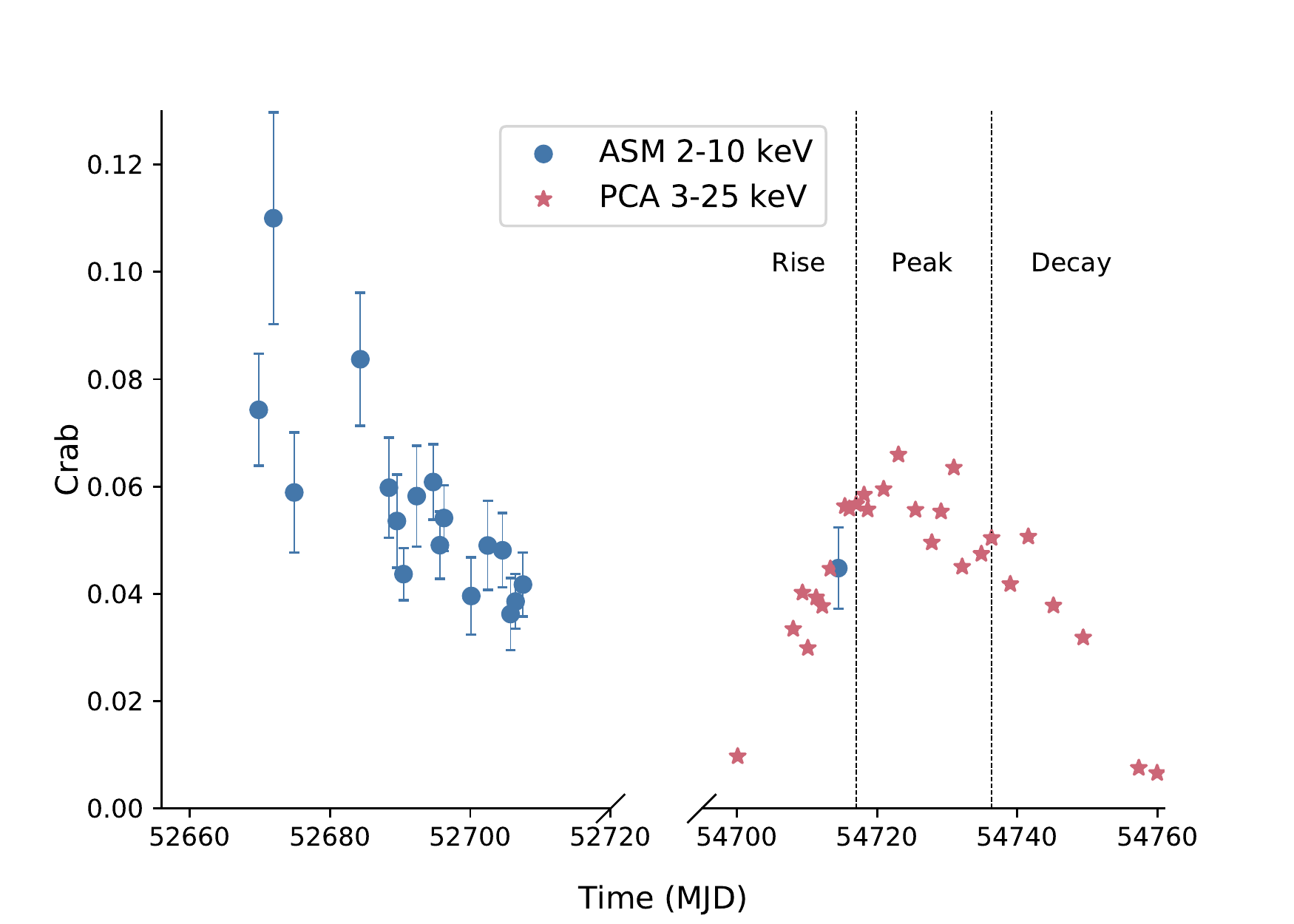}
    \caption{Persistent flux observations of XMMU J181227.8--181234 for the 2003 and 2008 outbursts from \textit{RXTE}. ASM is all sky monitor dwell-by-dwell sum band intensity (2-10 keV) with a 5$\sigma$ cut applied, and PCA is the proportional counter array mean flux (3-25 keV). Observations have been normalised to the Crab flux for the relevant instruments, assuming a Crab like spectrum. The rise, peak, and decay of the outburst as used in Figure \ref{fig:colourplot} are indicated.}
    \label{fig:persflux}
\end{figure}

When XMMU J181227.8--181234 was first detected in outburst by \textit{RXTE} on 2008-08-21 (MJD 54699) \citep{Markwardt2008}, the flux rose over 24 days to reach a peak on 54723, before dropping and returning to quiescent levels on 2008-10-19 (MJD 54758). The outburst lasted a total of approximately 60 days, similar to the 2003 outburst that lasted 60-100 days \citep{cackett2006}. The persistent flux observations for both the outbursts are plotted in Figure \ref{fig:persflux}. The conversion factors from count s$^{-1}$ to Crab for each telescope were used as follows: for ASM data 1 Crab = 75 count s$^{-1}$ and for PCA data 1 Crab = 2000 count s$^{-1}$ PCU$^{-1}$. The reader may note that there is some uncertainty associated with these conversion factors due to the variable nature of the Crab Nebula (variations of up to 7$\%$; \citealp[e.g.][]{Hodge2011}) and different telescope calibrations. The source was not detected by IBIS/ISGRI observations from the \textit{INTEGRAL} Galactic bulge monitoring program \citep{kuulkers2007}, with a typical 3$\sigma$ upper limit of 6-10$\,$mCrab in 18-40$\,$keV for each one of 69 3.5$\,$hr observations during the outburst.

We measured a peak persistent intensity of 97$\pm$3$\,$c s$^{-1}$ PCU$^{-1}$ in the 2-25$\,$keV energy band from the PCA, which translates to approximately 65$\,$mCrab (or 1.3$\times$10$^{-9}\,$erg s$^{-1}$cm$^{-2}$ for a Crab-like spectrum). For this outburst we adopted a bolometric correction ($c_{\rm bol}$) of 2.522, calculated based on the assumed $N_{\rm H}$ of 12.8$\times$10$^{22}\,$cm$^{-2}$ \citep{cackett2006}, and a Comptonisation model fit to the spectrum, outlined below. Spectral variations outside of the PCA band might contribute errors of up to $\approx 40\%$ \citep[e.g.][]{Thompson2008} to the bolometric correction, but these cannot be measured and are not taken into account.

We chose 3 observations near the peak of the outburst (Obs IDs 93044-11-04-01, 93044-11-04-02, 93044-11-05-00) to analyse for a high signal-to-noise persistent spectrum. We fit the three spectra simultaneously with two choices for the continuum model, first a blackbody and powerlaw, and second a Comptonisation model \cite[{\tt compTT} in {\sc XSpec};][]{tit94}. Only the normalisation was allowed to vary between the three observations, to account for variations in the accretion rate between the different epochs.

In both cases we included a multiplicative component to model the effects of neutral absorption along the line of sight using the absorption model from \citet{Morrison1983} (model `tbabs' in {\sc XSpec}). Hydrogen column density was fixed at $N_{\rm H}=12.8\times10^{22}\ {\rm cm^{-2}}$ \citep{cackett2006}; and a Gaussian component was included to account for an excess around 6.4~keV, likely arising from Fe~K$\alpha$ emission local to the binary or across the FOV. We also adopted a systematic error of 0.5\% channel to channel as recommended by the PCA instrument team \citep{shaposhnikov2012}. 

The persistent spectrum was equally well fit with the Comptonisation model as a blackbody plus power-law model. The persistent spectrum was soft, with a plasma temperature of kT = 2.73$\pm$0.05$\,$keV for a Comptonisation model with a Gaussian component, with a reduced chi-squared value of 1.00 for 111 degrees of freedom (DOF). The Gaussian  component was centered at 6.63 keV with a 1-$\sigma$ width of 1.1$\,$keV, most likely due to an iron emission line in the spectrum. While a Comptonisation model is a more physically motivated fit to the data, we also found that a blackbody plus powerlaw model with a Gaussian component fits nearly equally as well to the 3 observations, with a reduced chi-squared value of 1.04 and a blackbody temperature of 1.78$\pm$0.02 keV. The Gaussian was fit at 6.5$\,$keV with a 1-$\sigma$ width of 0.3$\,$keV, again attributed to an iron emission line from somewhere in the field of view. The model fit parameters are listed in Table \ref{tab:spectrafits}. The quoted error for the persistent unabsorbed flux is statistical, including the contribution from uncertainty in the absorption column, but does not include systematic contributions from the absorption model, bolometric correction, continuum model, or absolute calibration.

\begin{table*}
\begin{minipage}{\textwidth}
	\centering
	\caption{Spectral fit parameters for the 2 different models used for 3 observations near the peak of the 2008 outburst of XMMU J181227.8--181234}
	\label{tab:spectrafits}
	\begin{tabular}{lccccccc}
		\hline
	 \multicolumn{7}{c}{Model 1: Blackbody + Gaussian + Powerlaw Model}\\
		\hline
		$N_{\rm H}$ (cm$^{-2}$) &Constant & kT (keV) & Gauss. line (keV) & Gauss. $\sigma$ (keV) & Powerlaw PhoIndex
		
		 & red. $\chi^2$ & Flux\footnote{Average unabsorbed 3-25 keV flux in  10$^{-9}$ergs/cm$^2$/s}\\

		12.8$\times 10^{22}$ & 1, 0.96, 0.94 & 1.78$\pm$0.02 & 6.50$\pm$0.08& 0.3$\pm$0.2& 3.03$\pm$0.03 & 1.04 for 111 D.O.F.
		& 1.69$\pm$0.04\\
		
		\hline
     \multicolumn{7}{c}{Model 2: Comptonisation Model + Gaussian}   \\
		\hline
		
		$N_{\rm H}$ (cm$^{-2}$) & Constant & T$_0$, kT (keV) & Gauss. line (keV) & Gauss. $\sigma$ (keV)& taup  & red. $\chi^2$ & Flux\\
		
		12.8$\times 10^{22}$ &1, 0.96, 0.94 & 0.4$\pm$0.3, 2.76$\pm$0.05 & 7.1$\pm$0.4 & 1.1$\pm$0.2& 5.2$\pm$0.2 & 0.96 for 111 D.O.F. & 1.71$\pm$0.06\\
		
		\hline
    \end{tabular}
\end{minipage}
\end{table*}

\begin{figure}
	\includegraphics[width=\columnwidth]{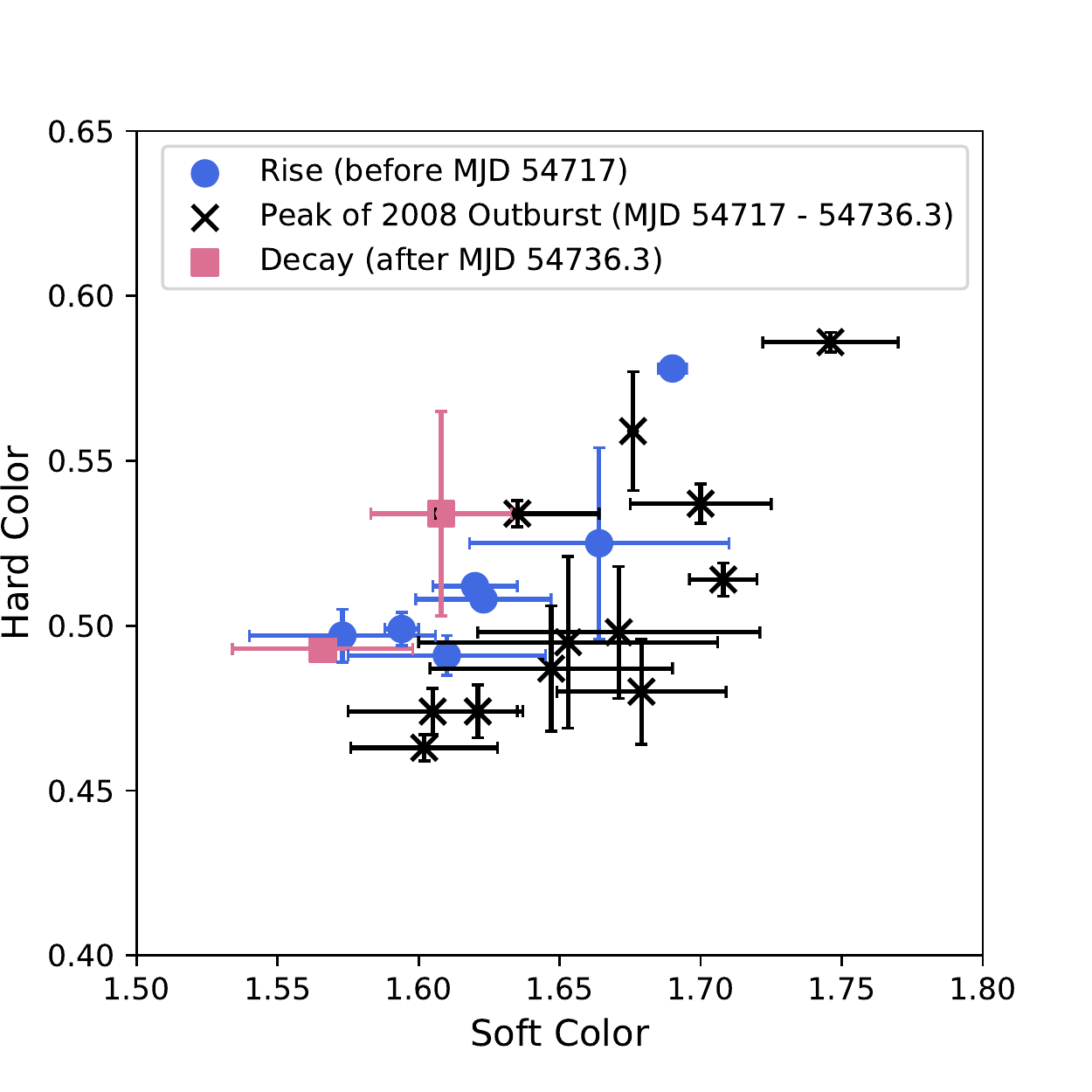}
    \caption{Evolution of the hard and soft colours of the source for the \textit{RXTE} PCA persistent flux observations. Observations begin at 54420.10115 and end at 55890.14174 so do not include the 2003 outburst of the source. Only detections are plotted, which are during the 2008 outburst. 
    There is no broad evolution of the colour during the different phases of the outburst for this source.}
    \label{fig:colourplot}
\end{figure}

We analysed the spectral evolution by looking at the soft and hard colours for the 2008 outburst. The soft and hard colours were found by following the approach used in \citep{galloway2008}, by calculating the ratios of fluxes in different bands rather than the usually adopted method for RXTE data that calculates the ratio of the counts in different bands. \citet{galloway2008} chose 4 bands: 2.2--3.6~keV, 3.6--5.0~keV, 5.0--8.6~keV and 8.6--18.6~keV, and calculated the soft and hard colour as the ratio of integrated fluxes in each pair of low and high energy fluxes. We found no significant evolution of the colour or temperature of the source over the observations, as seen in Figure \ref{fig:colourplot}, and could not deduce if the source is an atoll source on the banana branch or a Z source \citep{hasinger1989}. The kT and taup values listed in Table 1, model 2 are consistent with a soft persistent spectrum, so if the source was an atoll source this would imply it is on the banana branch.

\subsection{Quiescent Behaviour}\label{sec:longtermflux}

We explored the possibility of low-level activity from XMMU J181227.8--181234 outside of its known outbursts in order to properly classify the source. We analysed archival \textit{RXTE}/ASM taken from 1996 to 2012, \textit{Swift}/BAT (2004--current) and \textit{BeppoSax}/WFC (1996--2002) long-term observations of XMMU J181227.8--181234. In none of the data did we find any clear evidence for a third outburst, with a typical 3-sigma upper limit of 12 mCrab over a 1-week time scale in the 3--20 keV bandpass  for the ASM and 24 mCrab on a 1-day time scale in the 15-50 keV bandpass for \textit{Swift}/BAT.
Furthermore, we combined all WFC data (exposure time 4.5 Msec) into single images for the mission duration in 2--5, 5--10, 10--30 keV and combinations thereof. No point source is apparent at the location of XMMU J181227.8--181234 in these combined data
with a 5-sigma upper limit of 2 mCrab in 2--30 keV. A 500~s \textit{Swift}/XRT observation (Obs ID 00043947001) from 2012 September 29 centered just 1' from the position of XMMU J181227.8--181234 did not yield any detection, with a 3-$\sigma$ count rate of 0.012 count s$^{-1}$, a factor of 340 times fainter than the count rate observed by \textit{Swift} during its last outburst. This confirms in the 0.2-10$\,$keV range it was not persistently active. There are many bright sources in the neighbourhood of XMMU J181227.8--181234 (it lies between GX 9+1 and GX 13+1), which causes a lot of noise in the light curve of the source. We thus concluded that the low level persistent flux we observed from the source is attributed to this, and is not as a result of persistent activity from XMMU J181227.8--181234. Therefore, we classify XMMU J181227.8--181234 as a transient X-ray binary with episodes of high luminosity outbursts.

\subsection{Type I X-ray Bursts}

We detected 7 X-ray bursts in the data of the 26 follow-up observations with the PCA. Figure~\ref{fig:burstcounts} shows the light curves of these bursts. As the instrumental vignetting factor for the off-axis burster XTE J1810-189 would give a peak intensity that is roughly consistent with the measured peak intensity, the bursts could in principle come from that source. However, XTE J1810-189 was shown by the PCA bulge scans to be already in an off state for three months (see \S~\ref{sec:data}) and this possibility can be safely excluded. This leaves only XMMU J181227.8-181234 as the possible origin of the bursts.

The follow-up data have a total exposure time of 45 ks. The light curves of the burst events show rise times shorter than 2$\,$s, durations above the noise of about 20$\,$s, e-folding decay times of about 5$\,$s and net peak intensities of up to 180$\pm{10}\,$c s$^{-1}$ PCU$^{-1}$.

\subsubsection{Search for Burst Oscillations}

We searched for burst oscillations in all of the bursts observed, using Science Event files (E-125us-64M-0-1s), which have a time resolution of 122\,$\mu$s and employ 64-channels to read out photon energies between 2 and 60 keV. For each burst, we extracted photon sequences in the 20-s intervals starting 4\,s before the burst onset. Photon sequences were binned matching the time resolution and the Fourier transform was taken in 1-s sliding windows shifted by 0.5\,s with respect to each other. Real and imaginary Leahy-normalized coefficients \citep{Leahy1983} were saved for the harmonics between 1 and 2001 Hz. 

Overall, we found no strong oscillation signal or clustering of
oscillation candidates at any particular frequency down to the
threshold Leahy-normalized power of 13.816.
This power was converted to the background-corrected upper limit on fractional amplitudes \citep{watts2005}:
\begin{equation}
\label{eq:framp}
A = \left(\frac{P_\mathrm{s}}{ N_\mathrm{m} }\right)^{1/2}\frac{N_\mathrm{m}}{N_\mathrm{m}-N_\mathrm{bkg}}
\end{equation}
where $P_\mathrm{s}$ is the Leahy-normalized power of signal in the absence of noise, $N_\mathrm{m}$ is the number of photons in the given time window and $N_\mathrm{bkg}$ is the number of background  photons in the same time window, estimated as an average of the count rate in a minute-long region prior to burst onset. For unknown $P_\mathrm{s}$, we used the median value of $P_\mathrm{s} = P_\mathrm{m}+1$  from the distribution of $P_\mathrm{s}$ given $P_\mathrm{m}$  \citep{Groth1975,Watts2012}, where $P_\mathrm{m}$ is the power spectrum of noise + signal. The best upper limits on the fractional amplitude (at burst peaks) range from 24\% to 43\%. This is higher than all burst oscillation amplitudes measured thus far with RXTE \citep[e.g.][]{Bilous2018}.

\subsubsection{Time-resolved Spectral Analysis}

\begin{figure}
	\includegraphics[width=\columnwidth]{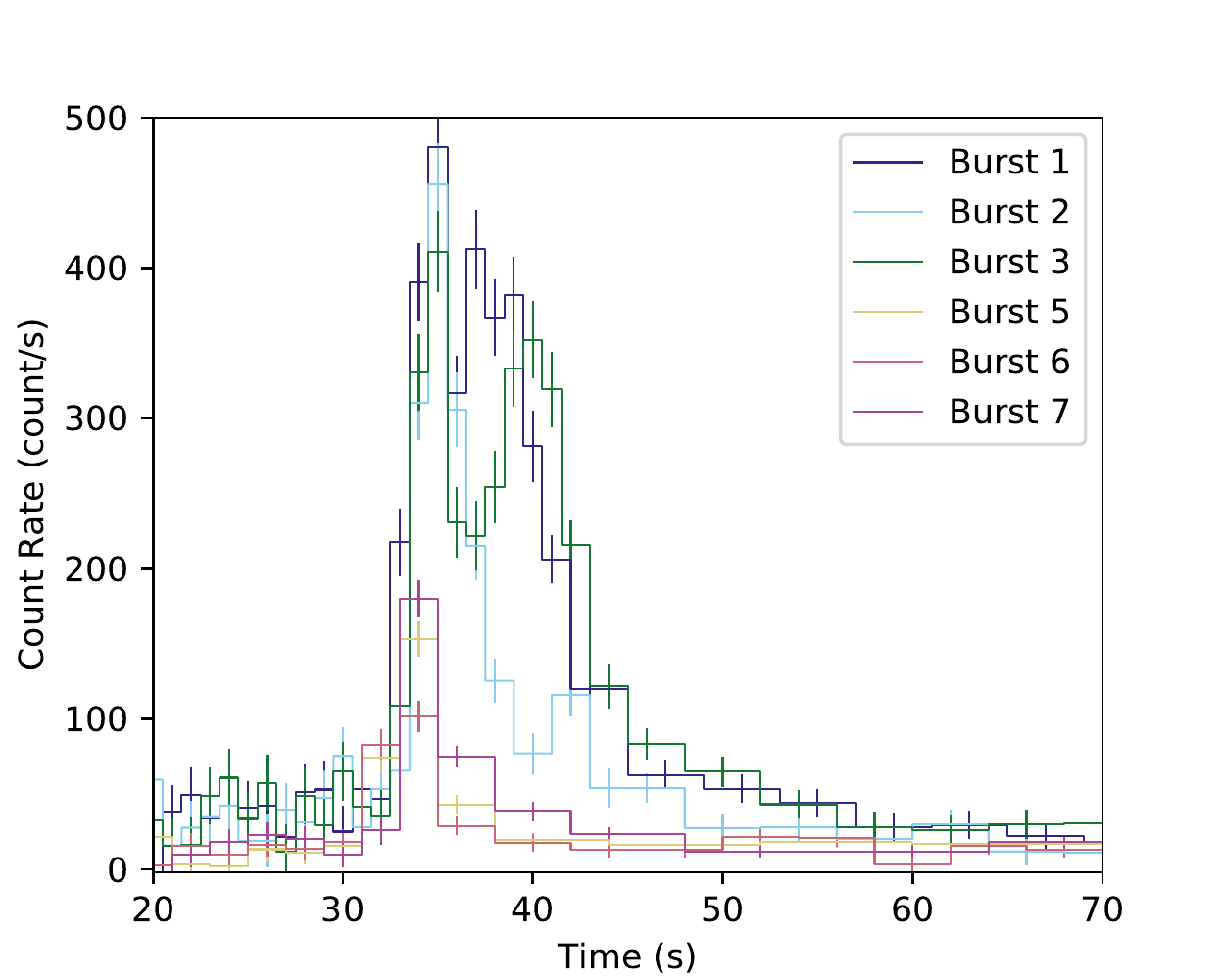}
    \caption{The light curves of 6 of the Type I X-ray Bursts observed during the 2008 outburst of XMMU J181227.8--181234. Observations are from the PCA aboard \textit{RXTE} and the onset of the burst was set to 33 seconds. The count rates have not been corrected for number of active PCUs. Bursts 5 and 6 are clear candidate SWT bursts, with significantly lower fluxes than the other observed bursts.}
    \label{fig:burstcounts}
\end{figure}

\begin{figure*}
	\includegraphics[width=\textwidth]{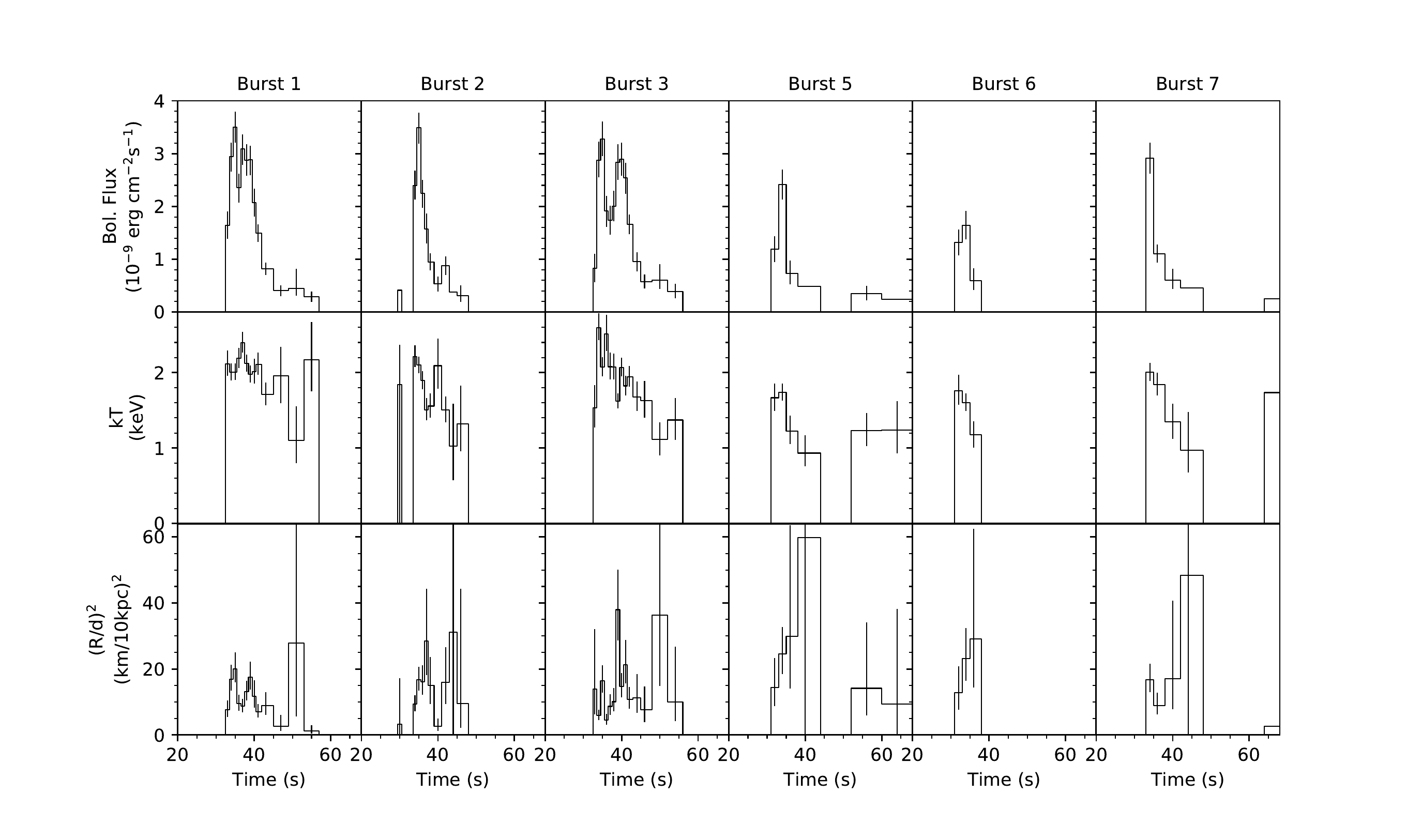}
    \caption{The time resolved spectroscopy of the four highest quality Type I X-ray bursts (bursts 1, 2, 3 and 7) and the two short waiting time bursts (bursts 5 and 6), from data taken with the the PCA aboard the \textit{RXTE}. Note the significant difference in flux of the SWT bursts compared to the normal bursts. There is signifcant evidence for cooling of the spectrum in all observed burst spectra, indicating these are Type I X-ray bursts. There is no evidence for radius expansion in any of the burst spectra.}
    \label{fig:burstspectra}
\end{figure*}

The burst spectra were modelled by black body radiation, slicing each burst in a number of time intervals to allow studying the spectral evolution. The number of time intervals varied from 3 to 20 for each burst, with the duration and number of the time intervals depending on the burst duration and data quality. The non-burst spectrum was assumed to be constant during the bursts as a first-order approximation since the data are of insufficient quality to probe changing accretion spectra during the burst \citep[e.g.,][]{worpel2013,zand2013,worpel2015}.
The results of this time-resolved spectral analysis of all bursts except burst 4, which does not have a high quality spectrum, are plotted in Figure \ref{fig:burstspectra}. 

We found the spectrum to be consistent with black body radiation, with a typical reduced $\chi^2$ value of 0.95. The peak temperature is kT = (2.4 $\pm{0.1}$) keV for burst 1 with a peak bolometric unabsorbed flux of (3.5 $\pm{0.3}) \times 10^{-9}\,$erg s$^{-1}$ cm$^{-2}$. 
We found significant evidence for cooling of the spectrum of the bursts, with the blackbody temperature decreasing from kT $\approx$ 2--2.5$\,$keV at the peak to kT $\approx$ 1$\,$keV after the peak for all bursts, as seen in Figure \ref{fig:burstspectra}. This confirms the bursts are Type I thermonuclear X-ray bursts. The radius variations of the bursts are not significant (< 3$\sigma$) and so we found no evidence for photospheric radius expansion in any of the bursts. There are distinct double peaks in the bolometric flux of bursts 1 and 3, with the second peak being slightly smaller in both cases. Similar double-peaked events have been observed in other burst sources without radius expansion, most notably 4U 1636$-$536 \citep[e.g.][]{Sztajno1985}.

\subsubsection{Burst Properties}

The observed and inferred burst properties are listed in Table \ref{tab:burst_properties}. We generally inferred burst properties based on the methods outlined in \citet{galloway2008}. $\alpha$ is the ratio of observed integrated persistent flux to observed burst fluence:
\begin{equation}\label{eq:alpha}
    \alpha = \frac{F_{\rm pers} c_{\rm bol} \Delta t}{E_{\rm b}}
\end{equation}
where $c_{\rm bol}$ is the bolometric correction factor, $F_{\rm {pers}}$ is the persistent flux, $\Delta t$ is the recurrence time and $E_{\rm b}$ is the burst fluence. Fluence is the observed bolometric integrated burst flux.

We estimated the average burst recurrence time using Poisson statistics by dividing the $45\,$ks (0.3446-day) exposure time by the number of the observed bursts, excluding burst 6, as we identified this as a candidate short waiting time (SWT) burst (see \S\ref{sec:swtbursts}). Since SWT bursts are caused by a different mechanism to `normal' bursts, we think it is appropriate to exclude the second SWT burst when inferring the recurrence time. We found the most likely recurrence time to be 1.4$^{+0.9}_{-0.5}\,$hours. We adopted a recurrence time of 18.0 $\pm$ 0.1 minutes for bursts 5 and 6 when calculating the $\alpha$ values. 

\begin{figure}
	\includegraphics[width=0.5\textwidth]{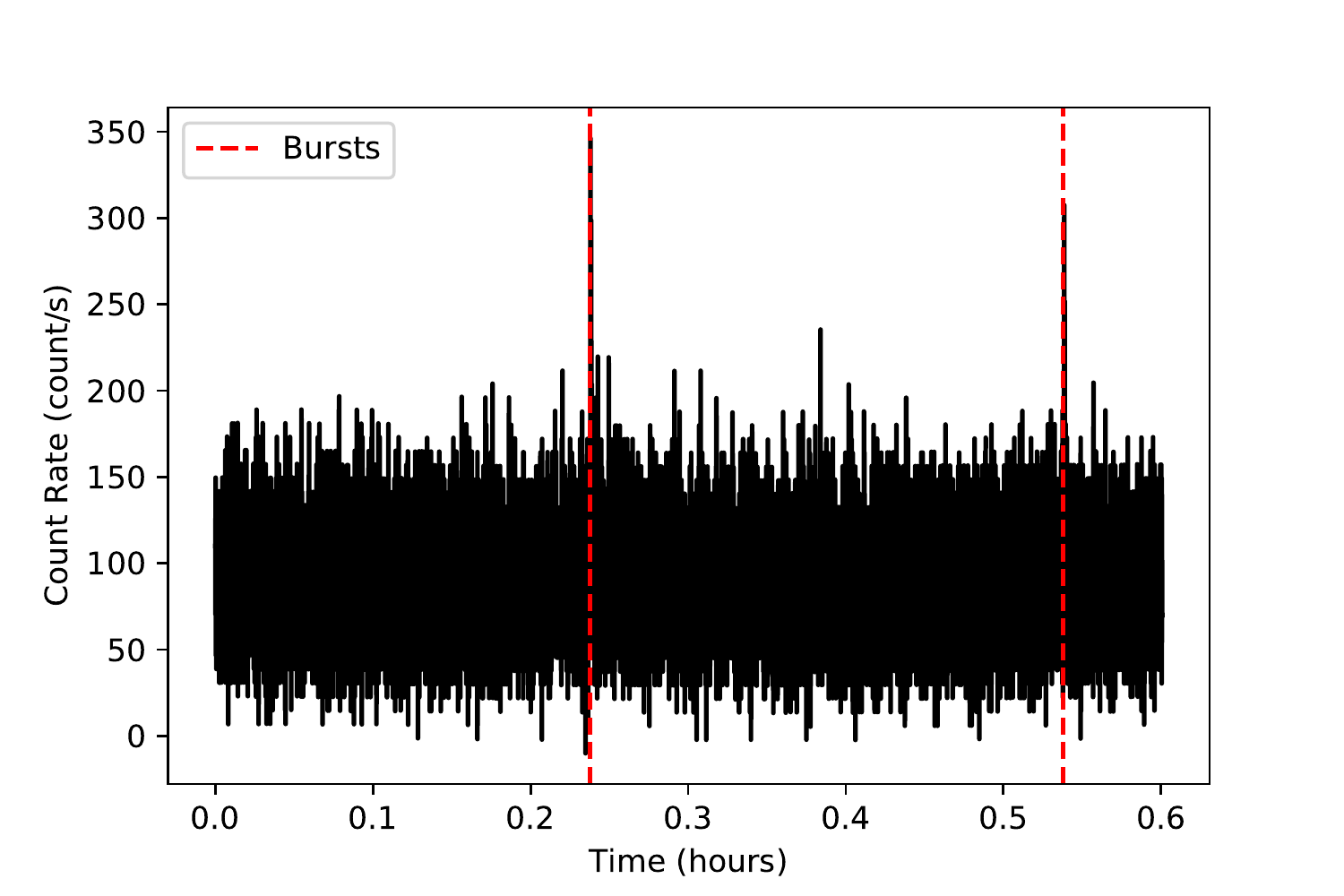}
    \caption{Raw \textit{RXTE} PCA lightcurve of the observation OBSID: 93044-11-05-02 in which the two SWT bursts (bursts 5 and 6) were observed. We cannot rule out SWT bursts occurring before or after these two bursts due to the length of the observation and the 0.3 hour recurrence time.}
    \label{fig:swtlc}
\end{figure}

We estimated the nuclear energy generation of the bursts, $Q_{\mathrm{nuc}}$ as 

\begin{equation}
\label{eq:qnuc}
    Q_{\mathrm{nuc}} = \frac{Q_{\mathrm{grav}}(1+z)}{\alpha}
     = \frac{218}{\alpha}\quad \mathrm{MeV/nucleon}
\end{equation}
where $1+z$ = 1.259 is the assumed gravitational redshift, $Q_{\mathrm{grav}}$ = $c^2 z/(1+z)$ $\approx$ GM$_{\rm NS}$/R$_{\rm NS}$,  is the gravitational energy, G is the gravitational constant and we use fiducial values of $M_{\rm NS} = 1.4\,M_{\odot}$ and $R_{\rm NS} = 11.2\,$km \citep[e.g.][]{Steiner2018} for the neutron star. 
The true value of $Q_{\rm nuc}$ thus depends linearly on the unknown neutron star mass, and
%
must also be corrected for possible anisotropy of the burst and persistent X-ray emission to find the true value in the NS frame (see Section \ref{sec:distance}). The anisotropy correction requires multiplication by  $\frac{\xi_b}{\xi_p}$, and explains why the $Q_{\rm nuc}$values reported in Table \ref{tab:burst_properties} are significantly less than the minimum expected value of 1.35~MeV~nucleon$^{-1}$.

\begin{table*}
\begin{minipage}{\textwidth}
	\centering
	\caption{Observed and inferred burst properties for the 7 Type I X-ray bursts detected during the 2008 outburst of XMMU J181227.8--181234.}
	\label{tab:burst_properties}
	\begin{tabular}{*{20}{lllllll}} 
		\hline
		Burst & 
		MJD &
		OBSID \footnote{Active PCUs: Bursts 1 and 2: PCU1 and PCU2, Burst 3: PCU2 and PCU3, Bursts 4, 5, 6 and 7: PCU2} &
		F$_{\rm p}$ \footnote{Peak bolometric burst flux. Note that we do not have flux measurements for burst 4, as the data quality was not good enough for this burst.}&
		E$_{\rm b}$ \footnote{Bolometric burst fluence. The values given in this table are the observed quantities and so are not corrected for anisotropy factors. See section \ref{sec:distance} for an explanation of the anisotropies.}&
		F$_{\rm pers}$ \footnote{3-25 keV persistent flux}&
		$\alpha$ \footnote{$\alpha$ is calculated between bursts using Equation \ref{eq:alpha}. For bursts 5 and 6, $\alpha$ was calculated assuming $\Delta t$ = 0.3$^{+0.002}_{-0.002}\,$hours and for all other bursts $\alpha$ was calculated assuming $\Delta t$ =1.4$^{+0.9}_{-0.5}\,$hours.}& 
		Q$_{\rm nuc}$\footnote{Inferred nuclear energy generation}&
		$\tau$\footnote{Decay time} \\

		& & & (10$^{-9}\,$ergs cm$^{-2}$)&(10$^{-9}\,$ergs cm$^{-2}$)&(10$^{-9}\,$ergs cm$^{-2}$s$^{-1}$)&&(MeV/nucleon)& (s)\\
		\hline

		1 & 54717.02685& 93044 11 03 02 & 3.5$\pm$0.3    & 35$\pm$2     & -  & -          &  -  & 5.0$\pm$0.4\\
		
		2 & 54718.13443  &  93044 11 03 03 &  3.5$\pm$0.3 & 20$\pm$5   & 1.16$\pm$0.03  &700$^{+500}_{-300}$   & 0.3$^{+0.2}_{-0.1}$  &4.5$\pm$0.7\\
		
		3 & 54720.94267 & 93044 11 03 05 &  3.3$\pm$0.3& 37$\pm$4   & 1.30$\pm$ 0.03        &400$^{+300}_{-200}$  &   0.5$^{+0.3}_{-0.2}$   &8.6$\pm$1.0\\
		
		4 & 54723.04537 &  93044 11 04 00 & -    & - & 1.45$\pm$0.03  &-   & - &-\\
		
		5 & 54732.12320  & 93044 11 05 02  &   2.4$\pm$0.3  & 15$\pm$2    &1.23$\pm$0.03           &230$^{+30}_{-30}$  &1.0$^{+0.1}_{-0.1}$   &1.3$\pm$0.2\\
		
		6 & 54732.13572  & 93044 11 05 02  &  1.6$\pm$0.3   & 8$\pm$1  &0.98$\pm$0.03   & 320$^{+40}_{-40}$  &0.7$^{+0.1}_{-0.1}$   &2.1$\pm$0.8\\
		
		7 & 54736.33460  & 93450 01 01 00  & 2.9$\pm$0.3   & 15$\pm$2    & 1.03$\pm$0.03 & 900$^{+600}_{-300}$  &0.3$^{+0.2}_{-0.1}$   &4.9$\pm$0.9\\
		\hline
    
	\end{tabular}
	

\end{minipage}
\end{table*}

\subsubsection{Short Recurrence Time Bursts}\label{sec:swtbursts}

Upon testing for regularity in the burst times we found that bursts 5 and 6 were just 18$\,$minutes apart, while the other bursts were observed days apart. We thus identified bursts 5 and 6 as candidates for short recurrence time bursts (or short waiting time (SWT) bursts). SWT bursts are Type I X-ray bursts that have short ($\lesssim$ 1$\,$hour) recurrence times and occur in groups of 2 or more (doubles, triplets or quadruplets) \citep[e.g.][]{linares2009,keek2010,keek2017,boirin2007}. \citet{keek2010} provided a comprehensive assessment of this phenomenon using a preliminary version of the MINBAR catalog of 3387 bursts observed with the \textit{BeppoSAX}/WFCs and \textit{RXTE}/PCA X-ray instruments. They found 136 SWT bursts with recurrence times of less than one hour, from 15 sources. Overall, they concluded that SWT bursts consistently have $\alpha$ values less than 40, lower peak luminosity, fluence, and temperature than `normal' bursts and lack the longer decay component from the rp-process in their decay profiles. Those authors also found that of the 15 sources found to exhibit SWT bursts, all were fast spinning ($>$500$\,$Hz) and of the sources where the accreted composition was known, they all had a high hydrogen fraction. SWT bursts have not been observed from ultra-compact X-ray binaries. 

Due to the low duty cycle of the observations of 1.78$\%$, we could not rule out burst 5 being the second or third burst in a triplet or quadruplet of SWT bursts and its lower fluence indicates this might be the case. Bursts 5 and 6 were observed in the same 0.6 hours telescope observation, with burst 5 occurring 0.24 hours from the start of the observation and burst 6 occurring 0.54 hours from the start of the observation. Since the average recurrence time of these bursts is 0.3 hours, we cannot rule out any bursts occurring after burst 6, as the observation continues for only 0.06 hours after burst 6 was observed. We also cannot rule out bursts occurring before burst 5, as the observation started 0.24 hours before burst 5 was observed, and the previous observation ended 1.147 days before this one began. We have plotted the raw lightcurve of the observation in which bursts 5 and 6 occurred in Figure \ref{fig:swtlc}. 

Bursts 5 and 6 in this outburst have $\alpha$ values significantly lower than the other bursts in the outburst, burst 6 has a short (18 minute) recurrence time, and we cannot rule out burst 5 also having a short recurrence time from the observations. The observations cannot constrain if these two bursts are part of a triplet or even quadruplet of SWT bursts. Both bursts have significantly shorter decay times than the other bursts in the outburst, which is typical of SWT bursts since they have been found to lack the longer decay component from the rp process in their decay profiles \citep{keek2010}.

The light curves of the two SWT bursts observed are plotted in Figure \ref{fig:burstcounts} and the time resolved spectroscopy of the bursts are plotted in Figure \ref{fig:burstspectra}. In these Figures it is clear that the SWT bursts (bursts 5 and 6) registered significantly lower fluence than the other bursts observed during this outburst, excluding burst 7, which we also cannot rule out as being an SWT burst from the duty cycle of the observations. The low fluence of bursts 5 and 6 provides strong evidence that these are SWT bursts. Bursts 5, 6 and 7 have fluences ranging from 8--15 $\times$10$^{-9}$ ergs cm$^{-2}$, on average a factor of 2.5 times smaller than the average fluence of the other bursts observed during the outburst, (20--37) $\times$10$^{-9}$ ergs cm$^{-2}$. However, we cannot confirm burst 7 as a SWT  burst as it has a decay time around the same as the non-SWT bursts in this outburst, and it is a combination of short recurrence time, low fluence, short decay time and low alpha values that characterises SWT bursts. The time resolved spectral analysis of bursts 5 and 6 showed significant evidence for cooling of the spectrum, confirming that these are also Type I X-ray bursts (Figure \ref{fig:burstspectra}).

\subsection{Distance, Accretion Rate and Anisotropy Factors}\label{sec:distance}



Here we estimate the system parameters based on the burst measurements described above.
The persistent and burst emission detected from LMXBs are both subject to projection effects due to the inclination angle of the accretion disk and its inclination-dependent obscuration of the NS. \citet{he2016} describe the angular distribution of radiation from low mass X-ray binaries for 4 different disk shapes. 
%
Anisotropy in the observed persistent emission is caused by scattering in the extended inner disk, collimating radiation from the boundary layer in the direction perpendicular to the disk plane. Anisotropy in the observed burst emission is caused by reflection of photons incident upon the disk surface, enhanced in the same direction as the persistent emission, as well as anisotropic emission from the burst itself. 
The simplest model (labeled `model A') assumes an optically thick, geometrically thin disk in the vicinity of the neutron star.

It is plausible that
this flat disk scenario is only applicable for sources in a soft spectral state, and a different disk geometry needs to be assumed for sources in a hard spectral state \citep[e.g.,][]{done2007}. Sources in the hard spectral state may instead have a truncated disk that would require different assumptions about the anisotropy factors. Since XMMU J181227.8--181234 remains in the soft state for the duration of this outburst (see Figure \ref{fig:colourplot}) 
we can safely use the calculations of \citet{he2016} to infer the anisotropy factors. 

We used the disk anisotropy model `A' geometry and the burst properties to place constraints on the inclination. 
From the recurrence time, average 
fluence (excluding the SWT candidate bursts 5 and 6) of (24$\pm{7}) \times 10^{-9}\,$erg cm$^{-2}$, 
and
persistent flux of (1.19$\pm{0.16}) \times 10^{-9}\,$erg cm$^{-2}$ s$^{-1}$ 
we calculated the average 
$\alpha$-value of the bursts as 
%
$620^{+280}_{-160}$.
%




We then drew samples from the $\alpha$ distribution and calculated the corresponding $Q_{\rm nuc}$ values from equation \ref{eq:qnuc}. 
For each sample we adopted a random value for the inclination $\theta$ drawn from an a priori isotropic distribution (i.e. uniform in $\cos \theta$). We calculated the anisotropy factors $\xi_b$, $\xi_p$, using the value for $\theta$ and the model of \citet{he2016}, and then estimated the H-fraction at ignition (adopting the 
%
$Q_{\rm nuc}$ value corrected for anisotropy) by inverting
the relation suggested by \citet{goodwin2018}:
\begin{equation}\label{eq:qnuckepler}
Q_{\mathrm{nuc}}\frac{\xi_b}{\xi_p} = 1.35 + 6.05\bar{X}\, \mathrm{MeV/nucleon}
\end{equation}
%
We retained only those samples with $\bar{X}>0$, excluding the unphysical values.

The H-fraction in the accreted fuel, $X_0$ can then be estimated based on the recurrence time and the estimated time to burn H prior to ignition \citep{lampe2016}:
\begin{equation}\label{eq:hfraction}
    t_{\mathrm{cno}} = 9.8 \,\mathrm{hr} \frac{X_0}{0.7}\frac{0.02}{Z}\\
\end{equation}
The implied H-fraction in the accreted fuel is $X_0 \leq 0.1071$ (95\% confidence), calculated using Equation \ref{eq:hfraction}. 

%
Only the smallest inclination values in the low tail of the $\alpha$-distribution satisfied our constraint on $\bar{X}$, indicating a likely range of  $\theta = 30\pm10$\textdegree\ (1-$\sigma$ confidence). Although only $\approx$3$\%$ of samples were retained, this fraction also depends on the CNO abundance. The fraction of samples accepted increases with lower metallicity, which might suggest that the metallicity in this source is lower than solar. Similarly, increased redshift also results in an increased acceptance fraction of samples. We note that this inclination is only correct under the assumption that the flat disk geometry assumed by \citet{he2016} Model A reflects the actual geometry of this system.  

%
Based on the possible range of inclination values derived above, we calculated the distance of XMMU J181227.8--181234. The accretion rate was inferred using the models of \citet{lampe2016} (Figure 1 in their paper) that relate recurrence time to accretion rate for different compositions. Given a recurrence time of 1.4 hours, the inferred $\dot{m}$ is $(0.30\pm{0.05})\,\dot{m}_{\rm Edd}$ \footnote{For simplicity we use $\dot{m}_{\rm Edd}$ = 8.8 $\times 10^{4}\,$g cm$^{-2}$ s$^{-1}$, but we note this Eddington accretion rate is assuming the hydrogen mass fraction of the fuel is X=0.7. In this paper this accretion rate is used merely as a reference point.} 
 for solar metallicity.
No photospheric radius expansion was detected so we found an upper limit for the distance of d$\sqrt{\xi_{\rm b}}$ = 25$\,$kpc, d = 35$\,$kpc for $\xi_b$ = 0.5, using the peak bolometric burst flux. Note the burst anisotropy factor inherent in the observations of the burst peak flux.

We inferred the distance using the anisotropy factors and Equation \ref{eq:distance}, modified from \citet{galloway2008}:

\begin{equation}
    \label{eq:distance}
    d^2 = \frac{G\dot{m} M_{\rm NS} R_{\rm NS}}{F_{\rm pers} \xi_{\rm p} c_{\rm bol} (1+z)} \mathrm{cm}^2
\end{equation}

\begin{figure}
	\includegraphics[width=\columnwidth]{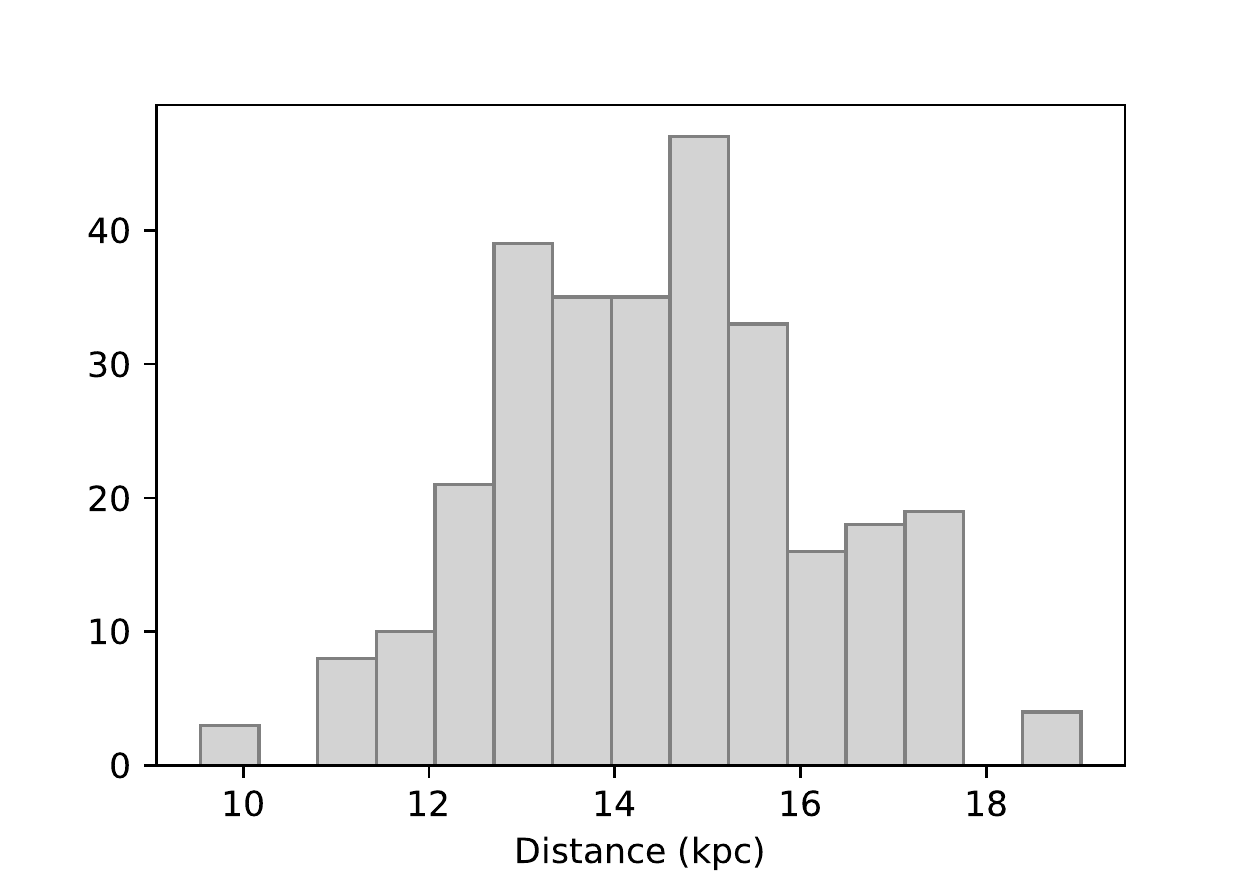}
    \caption{The distribution of distances to XMMU J181227.8--18123  calculated using Equation \ref{eq:distance} based on the allowed anisotropy factors. The $y$-axis corresponds to 
    the number of samples meeting our selection criteria (see Section \ref{sec:distance}).}
    \label{fig:distance}
\end{figure}

We found a distance to the source given by the distribution in Figure \ref{fig:distance}, with a median of 14.4 kpc and 68th percentile limits of +1.8 -1.6 kpc, for an accretion rate of 0.30$\pm$0.05$\,\dot{m}_{\rm Edd}$. Note that this distance estimate is dependent on the assumption that the NS mass is 1.4$\,$M$_{\odot}$ and radius is 11.2$\,$km. Recent works such as \citet{Antoniadis2016} or \citet{Steiner2018} have attempted to constrain the NS mass and radius for X-ray binary systems as well as millisecond pulsars. \citet{Antoniadis2016} found evidence that the birth mass of the NS in these systems could vary significantly, indicating that the NS constraints are highly system dependent, so the mass of the NS in XMMU J181227.8--181234 may not be well constrained by their, or other's observations. However, they find strong indication of a maximum NS mass of 2.018$\,$M$_{\odot}$. This corresponds to a 40$\%$ more massive NS than we have assumed, which would increase our distance estimate by up to 20$\%$. Likewise, for a minimum NS mass of 1.1$\,$M$_{\odot}$ \citep[e.g.][]{Antoniadis2016}, which is a 20$\%$ less massive NS, there would be a 10$\%$ decrease in the distance (which is within the statistical uncertainty of 15\%). \citet{Steiner2018} analysed observations of low mass X-ray binaries in quiescence and found that the NS radius for a 1.4$\,$M$_{\odot}$ is most likely between 10 and 14 km. These limits correspond to a 25\% increase or 10\% decrease in the radius, which would cause a 12.5\% increase or 5\% decrease in the distance. These radius effects are within the range of the statistical uncertainty on the distance of 15\%. Similarly, the bolometric correction could vary by up to 40$\%$ due to systematic effects (see Section \ref{sec:analysis}), which would cause a 20$\%$ change in the predicted distance. These systematic effects cannot be quantified and so we have not included them in the statistical uncertainty on the distance. We thus conclude the distance to XMMU J181227.8--181234 is 14$\pm$2 kpc.

A distance of 16$\,$kpc agrees with the typical distance of a Type I X-ray burst with peak luminosity of about 10$^{38}\,$erg s$^{-1}$ \citep{kuulkers2003} for the peak bolometric burst flux for this source of (3.5$\pm{0.3}) \times 10^{-9}\,$erg s$^{-1}$cm$^{-2}$. This is within 1-$\sigma$ of the distance we predict.

\section{Discussion and Conclusions}\label{sec:classification}

The discovery of Type I thermonuclear X-ray bursts observed from XMMU J181227.8--181234 confirms that the compact object is a neutron star in this system. The bursts have relatively short recurrence times and short decay times that indicate they lack the extended tail that results from rp-process burning in bursts with hydrogen. They also have low nuclear energy generation, low $\bar{X}$ and high $\alpha$ values, which all are properties found in (pure) helium bursts. Therefore, we deduce that the bursts ignite with almost pure helium fuel as the source of the thermonuclear burning. The short recurrence time requires the accreted fuel to be almost pure helium ($X_0 \leq 0.082$) as there is not sufficient time between bursts for a significant amount of hydrogen to burn to helium. 

We also observed several candidate short waiting time bursts. SWT bursts are usually observed in systems with high hydrogen fraction, so we have discovered SWT bursts with helium fuel for the first time. We tentatively classify XMMU J181227.8--181234 as an ultracompact X-ray binary due to the strong evidence that it is accreting almost pure helium fuel, despite the fact that SWT bursts have not been observed in ultracompact systems. We thus deduce that the companion star in the system is extremely evolved. We note that this kind of inference of an ultracompact nature (frequent short bursts at high accretion rates) is novel and adds to the methods of orbital period measurement, relatively faint optical counterparts and persistent accretion at very low levels \citep[$\la$1\% of Eddington, e.g.][]{zand2007,vanhaaften2012}. 

XMMU J181227.8--181234 has previously been tentatively classified as a transient high-luminosity distant low-mass X-ray binary \citep{cackett2006}. To further understand the behaviour of XMMU J181227.8--181234 we searched for some parallels with other known systems. We found 4 known helium bursters in the burst catalogue of \citet{galloway2008} with frequent, short bursts and a moderate to high accretion rate. These sources include Ser X-1, 4U 1728-34, 4U 1702-429 and 4U 1916-053. 4U 1728-34 and 4U 1916-053 are ultracompact systems, like XMMU J181227.8--181234. 4U 1916-053 and 4U 1702-429 are persistently active at low accretion rates, unlike XMMU J181227.8--181234. 4U 1728-34, 4U 1702-429 and 4U 1916-053 all have observed alpha values < 150, making them unlike XMMU J181227.8--181234, which has alpha values of 450-850. Ser X-1 is the only other source to exhibit high alpha values (as high as 1600), but this source also has irregular recurrence time bursts. SWT bursts have not been observed from any of these sources. Seemingly, none of the similar sources we found exactly match the behaviour of XMMU J181227.8--181234, making it an unusual source.

\citet{Keek2012} modelled superbursts in a helium accretor at $\approx$ 30$\%$ Eddington. After a superburst, the helium bursts evolve as the atmosphere cools. In the absence of a superburst and during a transient outburst, as in XMMU J181227.8--181234, similar bursting regimes may appear. Interestingly, the predicted recurrence times of the helium accreting model from \citet{Keek2012} exhibited two bursting modes: one with shorter and one with longer recurrence times. The longer recurrence time is similar to the average recurrence time of the bursts from XMMU J181227.8--181234, 80 $\pm$ 30 minutes. The peak luminosity of the modelled bursts is around 5$\times$10$^{38}$ erg s$^{-1}$, larger than the observed peak flux of the normal bursts from XMMU J181227.8--181234, but on the same order of magnitude. The shorter recurrence times are slightly longer than the 18 minutes of the SWT bursts from XMMU J181227.8--181234, at $\approx$45 minutes, and the peak flux is significantly smaller at $\approx$4$\times$10$^{37}$ erg s$^{-1}$. Nevertheless, this kind of bursting regime is similar to what we have observed in XMMU J181227.8--181234 and future work could tune the model parameters to replicate this kind of behaviour, and explore further what kind of bursting regimes are possible when accreting helium at high rates. 

XMMU J181227.8--181234 is a very frequent burster and so should be actively monitored at a higher duty cycle in its next outburst (perhaps within the next couple of years) as it will likely produce a large number of X-ray bursts for analysis.

\section*{Acknowledgements}

This paper utilizes preliminary analysis results from the Multi-INstrument Burst ARchive (MINBAR), which is supported under the Australian Academy of Science's Scientific Visits to Europe program, and the Australian Research Council's Discovery Projects (project DP0880369) and Future Fellowship (project FT0991598) schemes. 
%
The MINBAR project has also received funding from the European Union's Horizon 2020 Programme under AHEAD project (grant agreement n. 654215).
AG acknowledges support by an Australian Government Research Training (RTP) Scholarship.
We thank the anonymous referee for their detailed comments on the manuscript. 




\bibliographystyle{mnras}
\bibliography{bibfile}







\bsp	
\label{lastpage}
\end{document}